\documentclass[a4paper]{article}

\usepackage{INTERSPEECH2019}
\usepackage{url}
\title{Analyzing Verbal and Nonverbal Features for Predicting Group Performance}
\name{Uliyana Kubasova, Gabriel Murray, McKenzie Braley}
\address{
  University of the Fraser Valley\\
  Abbotsford, BC, Canada}
\email{uliyana.kubasova@student.ufv.ca, gabriel.murray@ufv.ca, mckenzie.braley@student.ufv.ca}

\begin{document}

\maketitle
\begin{abstract}
This work analyzes the efficacy of verbal and nonverbal features of group conversation for the task of automatic prediction of group task performance. We describe a new publicly available survival task dataset that was collected and annotated to facilitate this prediction task. In these experiments, the new dataset is merged with an existing survival task dataset, allowing us to compare feature sets on a much larger amount of data than has been used in recent related work. This work is also distinct from related research on social signal processing (SSP) in that we compare verbal and nonverbal features, whereas SSP is almost exclusively concerned with nonverbal aspects of social interaction. A key finding is that nonverbal features from the speech signal are extremely effective for this task, even on their own. However, the most effective individual features are verbal features, and we highlight the most important ones.
 
\end{abstract}


\section{Introduction}
Automated prediction of a group's performance on a task is useful for a number of reasons. For instance, a virtual meeting assistant with this capability might provide feedback to the group to enhance their performance and efficiency. As a second example, inspection of the trained model and its predictions could help a manager or team coordinator to better understand the aspects of group discussion that help or hinder performance. Because language is shown to relate to a group's performance on a task, with certain verbal and nonverbal cues facilitating greater performance \cite{fischer2007linguistic, reitter2014alignment, gonzales2010language, martin2004automated}, language is likely an important feature to consider when predicting group performance. Here, we address the challenge of automatically predicting performance on a winter survival task, where group members collaboratively rank a list of items in terms of their importance in a hypothetical survival scenario. Specifically, we automatically predict group scores using verbal and nonverbal features of group conversations, and compare the efficacy of these feature classes. The verbal cues we consider include linguistic, psycholinguistic, and syntactic cues, while the nonverbal cues we consider are acoustic features extracted from the speech signal. We use both the Group Affect and Performance (GAP) corpus \cite{braley2018group}, a survival task dataset which we have recently recorded and annotated, as well as an already existing survival task dataset, the Emergent Leadership (ELEA) corpus \cite{sanchez2012nonverbal}.

Contributions of this work are twofold. First, we release the GAP corpus with the goals of spurring research on the automated prediction of group performance and other related dynamics. Second, we compare verbal and nonverbal features for predicting group performance using a significantly larger dataset than what has been typically used. We achieve this increase in data by combining both the GAP and ELEA corpora. Previous work has either focused solely on nonverbal features \cite{avci2016predicting} or used a much smaller dataset to compare verbal and nonverbal features \cite{murray2018predicting}. Our main finding is that nonverbal features are very predictive of group performance on their own, though the most effective individual features are verbal. 

This trend of research is important, especially when considering that modern society has become increasingly complex and fast-paced, necessitating the use of teams to complete many operations. With this trend comes a need to improve the efficiency and quality of group work. Automated prediction of group performance can play a vital role here, by aiding our understanding of what enhances group performance, and providing real-time feedback to groups during their conversations.
This has been tested in a number of studies, although with mixed results \cite{tausczik2013improving, leshed2009automated, nowak2012social, dimicco2007impact}. 

In Section \ref{sec:related}, we discuss related work on predicting group task performance. In Section \ref{sec:features}, we describe the verbal and nonverbal features used in these experiments. Section \ref{sec:setup} includes the experimental setup, including description of the new corpus. Results are presented in Section \ref{sec:results}, and we conclude in Section \ref{sec:conclusion}. 

\section{Related Work}
\label{sec:related}
Automated prediction of group performance has been of increasing interest to the Social Signal Processing (SSP) community in recent years. SSP is concerned primarily with the nonverbal aspects of social interactions, including gesture, gaze, and prosody (see \cite{vinciarelli2012bridging} for a review). For example, Avci and Aran \cite{avci2016predicting} train models to predict group performance using the ELEA corpus. They use a variety of features, including personality traits derived from questionnaires, individual performance on the ranking task, visual gaze, and individual speaking cues such as speaking length and interruptions. The first key difference between our work and theirs is that in addition to nonverbal cues, we also use verbal cues, and compare these features for the task. A second key difference is that we only derive features from the conversation itself. That is, we do not use personality trait data from questionnaires, nor do we use individual task scores. Our aim is to determine if we can predict group performance based purely on analysis of the conversation.

More recently, Murray and Oertel \cite{murray2018predicting} also used the ELEA corpus to predict group performance. They compared verbal and nonverbal features for this task using the English subset of the corpus. We build off of this work by using a dataset that is twice as large, allowing for a broader comparison of the efficacy of the feature classes. Secondly, because their dataset was smaller, their work focused primarily on using data augmentation and domain adaptation to improve predictive performance. In contrast, we focus on gathering and annotating the GAP corpus data and supplementing that with the English subset of the ELEA corpus. 

Survival tasks have been a popular test-bed for studying social interaction, including computational methods for group interaction. For example, Sanchez et al. \cite{sanchez2012nonverbal} used the ELEA corpus to study how leadership emerges in group interactions where participants have no assigned roles. More recently, Okada et al. \cite{okada2019modeling} used the ELEA corpus and a job interview corpus for automatic prediction of personality trait and leadership impressions, using nonverbal features. Beyan et al. \cite{beyan2016detecting} have also addressed the task of automatically detecting emergent leaders, using an Italian-language survival corpus that is publicly available. 

Finally, there is related research using the Augmented Multimodal Interaction (AMI) meeting corpus \cite{carletta2007unleashing}. For example, Lai and Murray \cite{lai2018predicting} used multimodal features to predict group affect and satisfaction, and found that a combination of lexical, acoustic, and turn-taking features yielded the best predictive performance. 

\section{Features}
\label{sec:features}
In the following two sections, we describe the nonverbal features and verbal features, respectively, that were used in this set of experiments. 

\subsection{Nonverbal Features}

We extracted a large number of acoustic features from the audio recordings in the GAP and ELEA corpora using the openSMILE toolkit\footnote{\url{https://www.audeering.com/opensmile/}}. Specifically, we used the openSMILE configuration file that was originally developed for the INTERSPEECH 2010 Paralinguistic Challenge. This extracts a set of 1,582 acoustic features from each group recording. Due to the large number of features and small number of observations, we selected only the standard deviation features from the original set, which resulted in a final set of 76 speech features. These include jitter, shimmer, mel-frequency cepstral coefficients (MFCCs), associated delta features, PCM loudness, F0 envelope, F0 contour, and voicing probability. This is the same set of nonverbal features used by Murray and Oertel \cite{murray2018predicting}.

\subsection{Verbal Features}

We extract the following linguistic features from the transcripts of the meetings in both the GAP and ELEA corpora:

\textbf{Dependency Parse Features:}  Sentences are parsed using spaCy's dependency parser\footnote{\url{https://spacy.io/}}, and from these parses we extract several features, including sparse bag-of-relations features, type-token ratios for dependency relations, the branching factor of the root of the dependency tree, and the maximum branching factor of any node in the dependency tree.

\textbf{Part-of-Speech Tags:}  We also use spaCy's part-of-speech tagger, and extract a sparse bag-of-tags representation for the most frequent tags, as well as the type-token ratio for tags. 

\textbf{Filled Pauses:} We extract the total number of \textit{filled pauses}, such as `uh' or `hmm.' 

\textbf{Psycholinguistic:}  We use several psycholinguistic features that originate with word-level scores. All words are scored for their concreteness, imageability, typical age of acquisition, and familiarity, using publicly available ratings\footnote{\url{http://websites.psychology.uwa.edu.au/school/MRCDatabase/uwa_mrc.htm}}. We also derive SUBTL scores for words, which indicate how frequently they are used in everyday life \cite{brysbaert2009moving}. For each of these psycholinguistic ratings, we calculate an average score over all the words in the conversation, omitting any words that do not have ratings. 

\textbf{Sentiment:} We use the SO-Cal (Semantic Orientation Calculator) sentiment lexicon \cite{taboada2011lexicon}, which associates positive and negative scores with sentiment-bearing words, indicating how positive or negative their sentiment typically is. We then sum the sentiment scores over all of the sentiment-bearing words in the conversation.

\textbf{GloVe Word Vectors:} Words are represented using GloVe vectors\footnote{\url{https://nlp.stanford.edu/projects/glove/}}, and the vectors are summed over sentences. We then create a document vector that is the average of the sentence vectors. The first five dimensions of the document vectors are used as features. 

\textbf{Lexical Cohesion:} We measure cohesion using the average cosine similarity of adjacent sentences in a document, with sentences representing using the summed GloVe vectors.

\textbf{Sentence and Document Length:}  We extract the average number of words per sentence, and average number of sentences per meeting.

This is the same as the set of verbal features used by Murray and Oertel \cite{murray2018predicting}, except that we omit the bag-of-words features, because their inclusion results in a number of features that is much higher than the number of observations in these experiments.

\section{Experimental Setup}
\label{sec:setup}
In this section, we first describe the new dataset that was collected and annotated, and then give an overview of the machine learning models and evaluation metrics that are used in these experiments. 

\subsection{Dataset}

In our experiments, we use both the GAP \cite{braley2018group} and ELEA \cite{sanchez2011audio, sanchez2012nonverbal} corpora. The GAP corpus consists of 28 meetings where in each meeting, two to four group members engage in a recorded decision-making task, called the Winter Survival Task (WST). In this task, participants imagine that they have been in a plane crash. They must rank a list of items in terms of their importance to their survival in the hypothetical crash scenario. Performance on the task is quantified by comparing participant responses to survival experts responses, allowing for an objective measure of decision-making performance. The group members first complete the task individually, and then complete a recorded group version of the task, where they collectively come up with one group ranking. Although we do not look at this data in this work, group members also fill out a post-task meeting questionnaire consisting of items related to satisfaction with the meeting (e.g., whether they felt that the group worked well together, used its time wisely, etc.). The meeting recordings have been manually segmented into individual dialogue acts, with each dialogue act then manually transcribed verbatim and annotated for sentiment (positive or negative) and decision-making (proposal, agreement, disagreement, and confirmation). However, for this analysis, we only look at the meeting transcripts, audio, and group task performance. A total of 84 participants make up 16 groups of three, six groups of two, and six groups of four. The corpus consists of 266.16 minutes of meeting recordings total, with each meeting recording lasting on average 9.50 minutes. See \cite{braley2018group} for more information on data collection and preparation, as well as characteristics of the first set of 13 group meetings that were released in the first phase of the project. The full set of 28 meetings is freely available, and includes audio, transcripts, questionnaires, group performance metrics, and sentence-level annotations of sentiment and decision-making processes\footnote{\url{https://sites.google.com/view/gap-corpus/home}}. All of the conversations in the GAP corpus are in English, and all participants were fluent speakers of English.

We also use the ELEA corpus as a secondary source of data. The ELEA corpus consists of 40 groups of three to four members who also collaborate to complete a group WST ranking. In creating the GAP corpus, we replicated most aspects of the ELEA corpus, with the aims of supplementing the available ELEA data. The ELEA and GAP corpora are therefore very similar in terms of meeting procedures. The ELEA corpus consists of 40 group meetings total, with approximately 600 minutes of meeting recordings, and average meeting recording lengths of 15 minutes. There are 148 participants, with 28 teams of four and 12 teams of three. We use a subset of 29 meetings in English for our analyses.

Merging the GAP and ELEA corpora gave us a combined dataset of 57 groups. Each of the recordings are described by 76 nonverbal and 102 verbal features. Features containing all zero or missing values were removed.

Using both datasets, we predict the group scores on the WST. The group score is calculated by summing the absolute differences between the group ranking and the expert ranking for each item. Here, lower scores reveal greater similarity to the expert ranking and thus better task performance. 

\subsection{Models and Metrics}

\begin{figure}[htp]
\centering
\includegraphics[width=7cm]{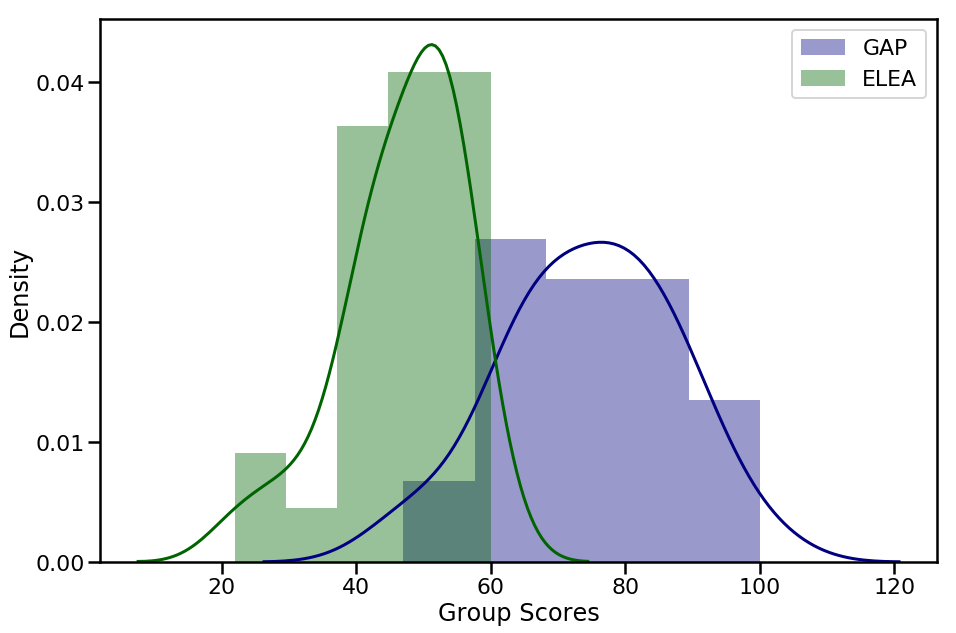}
\caption{\textbf{GAP and ELEA Group Scores Before Scaling}}
\label{fig:before}
\end{figure}

\begin{figure}[htp]
\centering
\includegraphics[width=7cm]{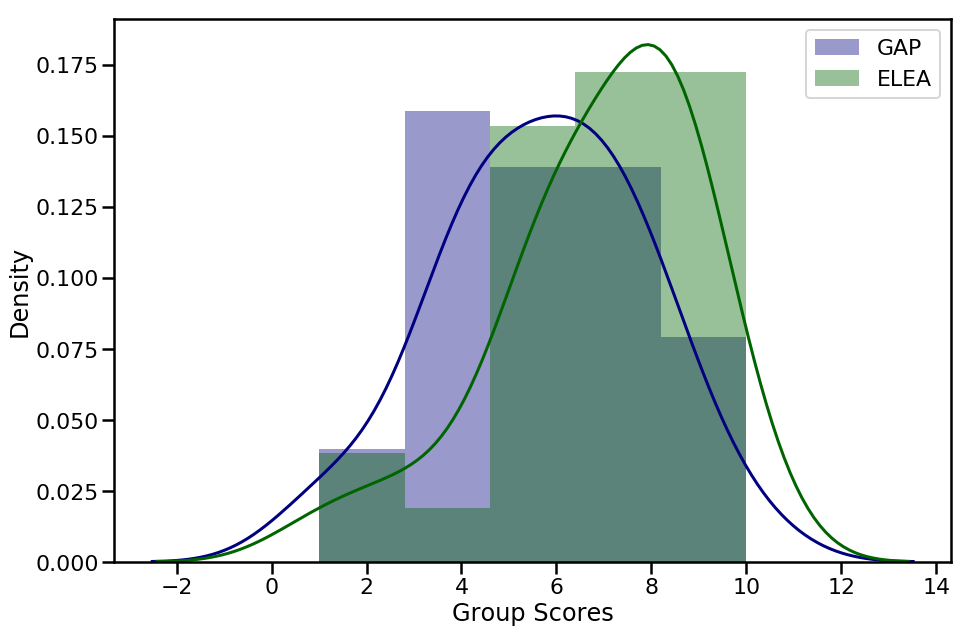}
\caption{\textbf{GAP and ELEA Group Scores After Scaling}}
\label{fig:after}
\end{figure}

The survival tasks used in the GAP and ELEA corpora are slightly different versions; the GAP corpus uses a WST with 15 items and the ELEA corpus uses a WST with 12 items. Because participants from the ELEA corpus rank a lower number of items, group scores in the ELEA corpus are lower. Specifically, the mean of the ELEA group scores is 46.90 with a standard deviation of 8.91, whereas the mean of the GAP group scores is 77.61 with a standard deviation of 12.45. In order to evaluate both corpora on the same scale, we transformed the scores by scaling them to a range between 1 and 10 using the MinMaxScaler. Figures \ref{fig:before} and \ref{fig:after} show the distribution of the scores before and after scaling was performed, respectively.

For these experiments, we compare three tree-based regression models: Random Forest (RF), Gradient Boosting (GB) and Extra Trees (ET). All models were trained using the Scikit-Learn Python package. We used the default training parameters except for the number of estimators, which was set at 50. 

Due to the small amount of data, we employ 5-fold cross-validation. 
The accuracy of our models is evaluated using mean squared error (MSE). We compare the performance of the three tree-based ensemble models by varying the set of features used for predictions and reporting the variable importance for the best performing models. There are three sets of experiments we are highlighting in this work: nonverbal features only, verbal features only, and verbal and nonverbal features combined. We compare all models with a mean prediction baseline, where the mean group score from the training folds is used as the prediction for the test fold observations.

\section{Results}
\label{sec:results}

Table \ref{tab:mse} shows the MSE scores for all models. The best-performing model in terms of MSE is using just nonverbal features with RF. According to paired t-tests, this best model gives a marginally significant improvement (0.05$<$\textit{p}$<$0.1) over the mean baseline, and a highly significant improvement  (\textit{p}$<$0.01) over the worst-performing model (all features with Gradient Boosted Trees). Furthermore, this best-performing model has an $R^2$ (goodness-of-fit) value of 0.8630, indicating that the nonverbal features are able to explain a large amount of the variance in the dataset.

\begin{table}[h]
\begin{tabular}{|l|c|c|c|} \hline
\textbf{Model} & \textbf{Nonverbal} & \textbf{Verbal} & \textbf{All Feas.} \\ \hline
Mean Baseline & \multicolumn{3}{|c|}{$\cdots$ \textbf{5.2974} $\cdots$ } \\\hline
Random Forests & \textbf{4.2511} & 5.8794 & 5.6673   \\
Gradient Boosted Trees & 5.3064 & 6.6434 & 6.8719  \\
Extra Trees & 4.6848 & 5.9286 & 6.0776 \\ \hline
\end{tabular}
\vspace{0.4cm}
\caption{Mean Squared Errors for All Models}
\label{tab:mse}
\end{table}

\begin{figure}[htp]
\centering
\includegraphics[width=8cm]{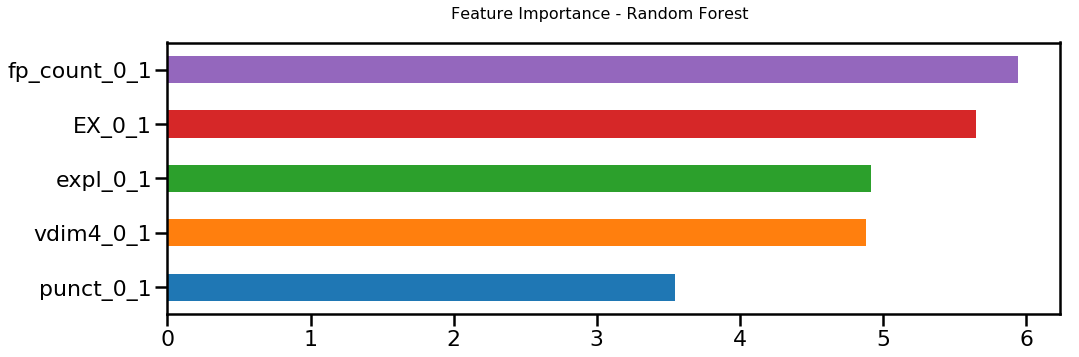}
\caption{\textbf{Key Overall Features w/ RF Models}}
\label{fig:allfeas}
\end{figure}

None of the models incorporating verbal features outperform the baseline. That being said, we also analyze individual features in terms of their importance, where importance is defined as the average reduction in MSE when the feature is used as a split point in a decision tree in the RF model. The most important individual features are verbal features, as shown in Figure \ref{fig:allfeas}. The top individual feature is the number of filled pauses (\textit{fp\_count\_0\_1}), followed by dependency relations, part-of-speech tags, and a feature derived from GloVe embeddings (\textit{vdim4\_0\_1}). This shows that although models using the broad verbal feature class are not effective in predicting group performance, single verbal features are indeed effective for predicting group scores.

\begin{figure}[htp]
\centering
\includegraphics[width=7cm]{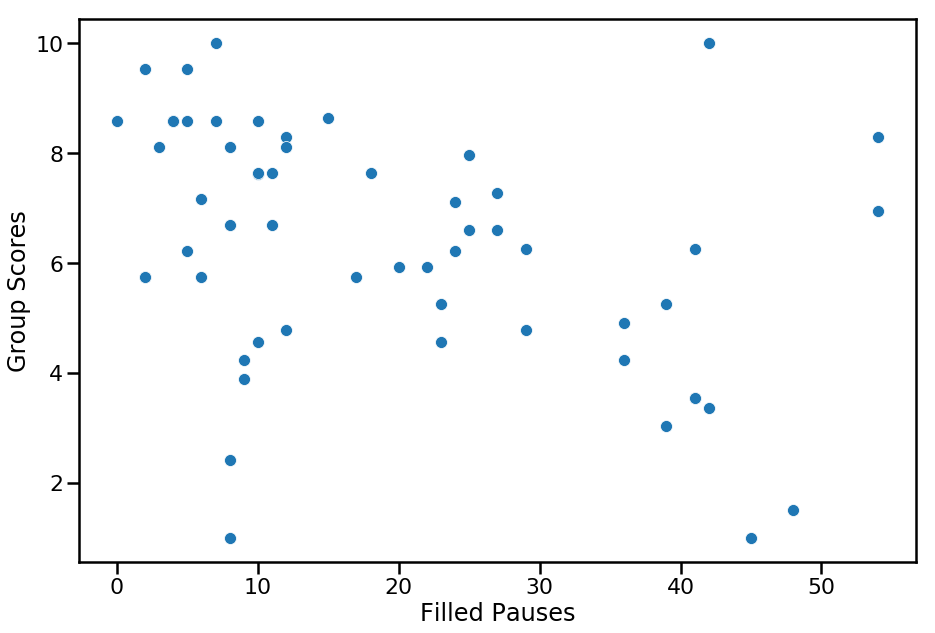}
\caption{\textbf{Filled Pauses vs. Group Scores}}
\label{fig:fp-scores}
\end{figure}

Figure \ref{fig:fp-scores} shows the relationship between the number of filled pauses and the group score, and there is a clear correlation, with a higher number of filled pauses being associated with a lower (i.e. \textit{better}) group score. While the number of filled pauses will tend to be larger for longer meetings, the filled pause feature is much more important than the separate meeting length feature. One hypothesis is that group members express uncertainty through these filled pauses, and that groups featuring members who are uncertain about their own rankings will more carefully deliberate and subsequently perform better. 

\begin{figure}[htp]
\centering
\includegraphics[width=8cm]{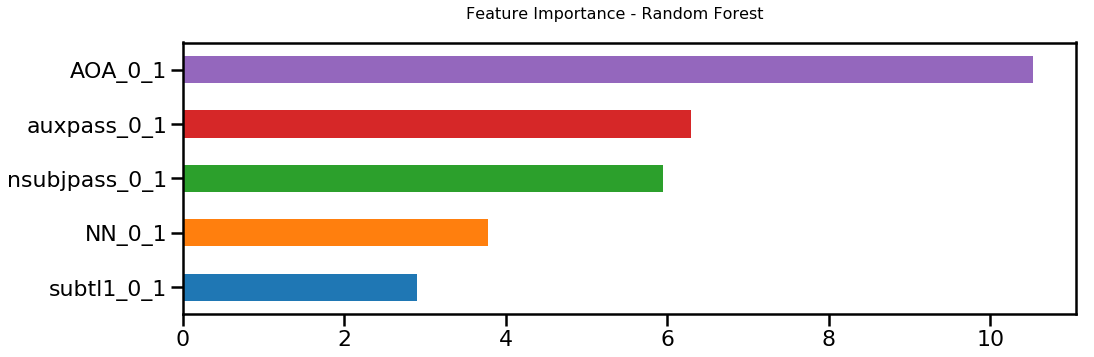}
\caption{\textbf{Key Verbal Features w/ RF Models}}
\label{fig:lingfeas}
\end{figure}

Figure \ref{fig:lingfeas} shows the most important features in the RF models when trained only with verbal features. Two features are psycholinguistic: age-of-acquisition (\textit{AOA\_0\_1}) and SUBTL score (\textit{subtl1\_0\_1}). The other three are again dependency relations and part-of-speech tags. 

\begin{figure}[htp]
\centering
\includegraphics[width=8cm]{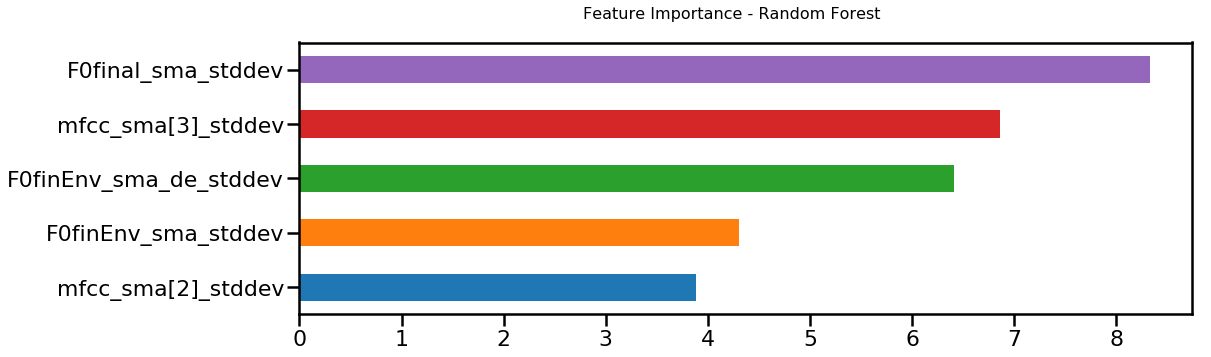}
\caption{\textbf{Key Nonverbal Features w/ RF Models}}
\label{fig:speechfeas}
\end{figure}

Figure \ref{fig:speechfeas} shows the most important features in the RF models when only training with nonverbal features. The top five are all F0 and MFCC features. 

Finally, given the small sample size and a large amount of features, we also tried feature selection in combination with RF regression, as well as reducing the number of features using Principal Components Analysis (PCA). However, these experiments did not yield any further improvement in prediction performance compared with the results presented above. Further experiments with neural models and lasso regression also gave much worse performance.

Our finding that verbal/linguistic features alone did not yield good prediction performance is, in fact, compatible with the findings of Murray and Oertel \cite{murray2018predicting}, who initially had the same result. They subsequently applied domain adaptation and data augmentation and found that the verbal model was amongst the best, second only to the combined verbal+nonverbal model. In lieu of doing domain adaptation or data augmentation, in this work we combined a newly gathered dataset with the existing ELEA corpus, effectively leading to a two-fold increase in the amount of data. However, the data augmentation method by Murray and Oertel \cite{murray2018predicting} increased the amount of training data four-fold. In future work, we will attempt to replicate the effectiveness of their data augmentation scheme on the merged dataset. Even with the bag-of-words features removed, many of the verbal features are sparse and will likely require additional data in order for that verbal model to be maximally effective.


\section{Conclusion}
\label{sec:conclusion}

In this work, we show that nonverbal and verbal features from a group conversation are predictive of the group's performance on a task. Specifically, nonverbal features from the speech signals are most effective as a class, although the most effective individual features are verbal. In contract with previous work that focuses primarily on nonverbal aspects conversations (e.g., \cite{avci2016predicting}), we also consider verbal aspects, showing that verbal features are also effective for predicting group performance and thus deserve an increased focus in the computing literature. 

We also use a dataset that is significantly larger than datasets previously used (e.g., \cite{murray2018predicting}). This is therefore the largest comparison of verbal and nonverbal features for the task of predicting group performance. We achieved this increase in data by creating a new dataset, the GAP corpus, and by combining the new data with the ELEA corpus. We also use this work to present the GAP corpus, which is being made available for research purposes. Because of the expensive and time-consuming nature of data collection and preparation, there exists a limited number of small groups corpora. It is our hope that recording, transcribing, annotating, and releasing the GAP corpus data will address this gap and stimulate automated analyses of small group performance and other related dynamics. 

In future work, we will attempt to further improve performance on this task through data augmentation and domain adaptation. We will also extract finer-grained verbal features that capture information about how language changes over the course of a conversation.

\textbf{Acknowledgements} The authors are supported by an NSERC Discovery Grant [RGPIN-2018-06806]. 

\bibliographystyle{IEEEtran}
\bibliography{mybib}

\end{document}